\documentclass[manuscript,nonacm]{acmart}

\usepackage[T1]{fontenc}
\usepackage{textcomp}
\usepackage{graphicx} %
\usepackage{ccicons}  %

\usepackage{subcaption}

\setcopyright{none}

\title{Content-Aware Automated Parameter Tuning for Approximate Color Transforms}

\author{Chatura Samarakoon}
\email{cts32@cam.ac.uk}
\orcid{1234-5678-9012}
\affiliation{%
  \institution{University of Cambridge}
  \streetaddress{Department of Engineering}
  \city{Cambridge}
  \postcode{CB3 0FA}
  \country{UK}
}

\author{Gehan Amaratunga}
\email{gaja1@cam.ac.uk}
\affiliation{%
  \institution{University of Cambridge}
  \streetaddress{Department of Engineering}
  \city{Cambridge}
  \postcode{CB3 0FA}
  \country{UK}
}

\author{Phillip Stanley-Marbell}
\email{ps751@eng.cam.ac.uk}
\affiliation{%
  \institution{University of Cambridge}
  \streetaddress{Department of Engineering}
  \city{Cambridge}
  \postcode{CB3 0FA}
  \country{UK}
}

\renewcommand{\shortauthors}{Samarakoon, et al.}

\begin{document}

\begin{abstract}
    There are numerous approximate color transforms reported in the literature that aim to reduce display power consumption by imperceptibly changing the color content of displayed images.
    To be practical, these techniques need to be content-aware in picking transformation parameters to preserve perceptual quality.
    This work presents a computationally-efficient method for calculating a parameter lower bound for approximate color transform parameters based on the content to be transformed.
    We conduct a user study with 62 participants and 6,400 image pair comparisons to derive the proposed solution.
    We use the user study results to predict this lower bound reliably with a 1.6\% mean squared error by using simple image-color-based heuristics.
    We show that these heuristics have Pearson and Spearman rank correlation coefficients greater than 0.7 (p<0.01) and that our model generalizes beyond the data from the user study.
    The user study results also show that the color transform is able to achieve up to 50\% power saving with most users reporting negligible visual impairment.

\end{abstract}

\maketitle

\section{Introduction}

The consumer demand for mobile devices capable of displaying 4K high definition video and content with high dynamic range (HDR) has shifted the mobile display market towards using emissive pixel display technologies like Organic Light Emitting Diodes (OLEDs) and Quantum Dot Light Emitting Diodes (QD-LEDs). However, the increase in pixel density and the display luminance requirements for HDR has caused these displays to consume an increasing proportion of overall device power~\cite{Chen2015a}. This trend will continue for the foreseeable future as the efficiency of displays are bounded by analogue optoelectronic processes that limit their ability to be miniaturized.

In contrast to traditional backlit displays, the power consumption of an emissive pixel display is a function of the content being displayed~\cite{4DSystems2008}. This presents new opportunities to optimize display power from software by modulating the content on the screen. \textit{Crayon}~\cite{Stanley-MarbellMIT2016}, transforms images to reduce display power consumption by imperceptibly changing the color and shape content of images displayed. Crayon was shown to achieve up to 60\% power saving while preserving image quality. However, the Crayon color transform had a parameter $\lambda$ that needed to be manually tuned to achieve the desired trade-off between display power savings and transformed image acceptability, limiting its broad acceptability.

\begin{figure}[h]
    \begin{minipage}{0.3\linewidth}
        \centering
        \includegraphics[width=0.9\linewidth]{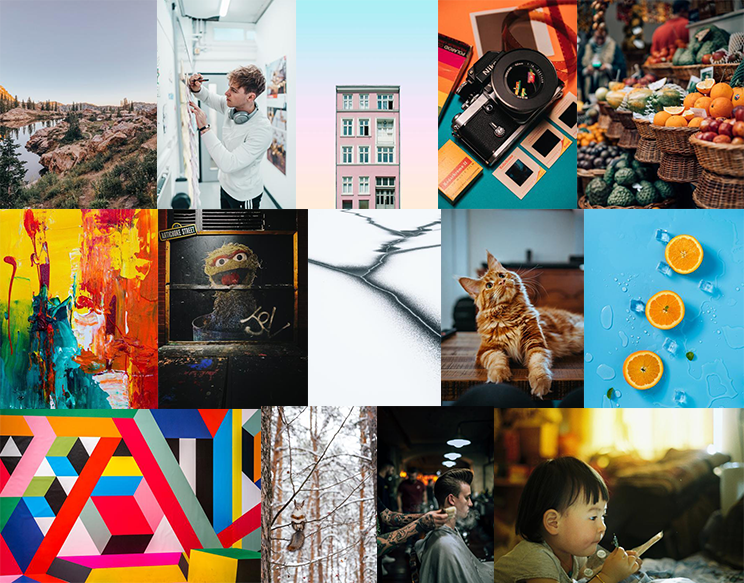}
        \caption{Collage of images used in the study to highlight the diversity of images.}
        \label{fig:imageCollage}
    \end{minipage}\hfill
    \begin{minipage}{0.35\linewidth}
        \centering
        \begin{subfigure}[t]{.15\linewidth}
            \centering
            \includegraphics[width=.99\linewidth]{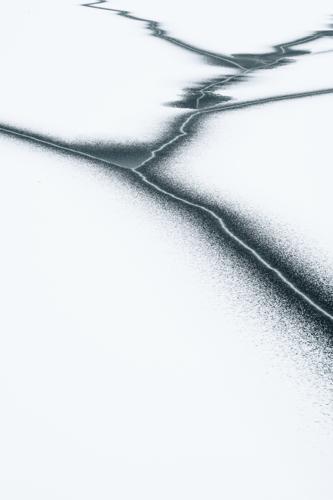}
            \caption{} %
        \end{subfigure}%
        \begin{subfigure}[t]{.15\linewidth}
            \centering
            \includegraphics[width=.99\linewidth]{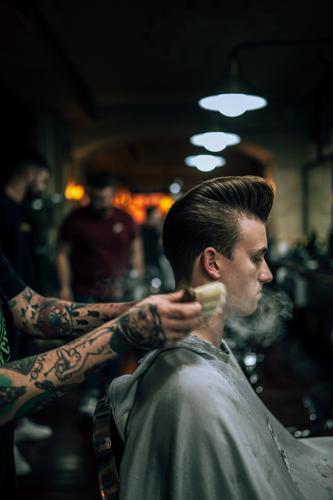}
            \caption{} %
            \label{fig:reis}
        \end{subfigure}%
        \begin{subfigure}[t]{.15\linewidth}
            \centering
            \includegraphics[width=.99\linewidth]{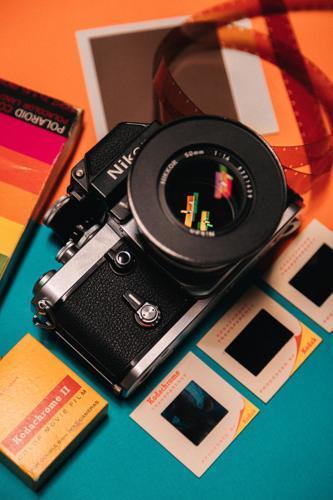}
            \caption{} %
        \end{subfigure}
        \begin{subfigure}[t]{.15\linewidth}
            \centering
            \includegraphics[width=.99\linewidth]{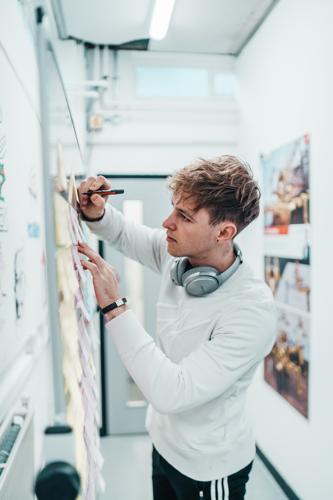}
            \caption{} %
        \end{subfigure}%
        \begin{subfigure}[t]{.18\linewidth}
            \centering
            \includegraphics[width=.99\linewidth]{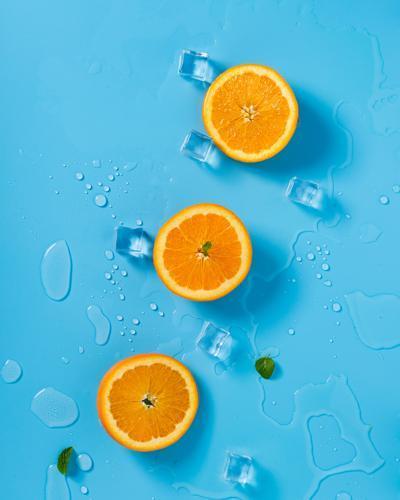}
            \caption{} %
        \end{subfigure}%
        \begin{subfigure}[t]{.18\linewidth}
            \centering
            \includegraphics[width=.99\linewidth]{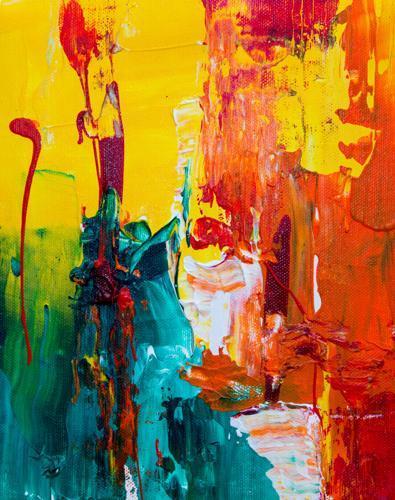}
            \caption{} %
        \end{subfigure}
        \caption{Only the results from the above subset of images are used in Figures \ref{fig:mosVlambda} and \ref{fig:lbEstCurves} for increased readability. The images are referenced by the author's last name; (a) Hojo, (b) Reis, (c) Owens, (d) Rajendharkumar, (e) Mu, (f) Johnson.}
        \label{fig:egImages}
    \end{minipage}\hfill
    \begin{minipage}{0.3\linewidth}
        \centering
        \begin{subfigure}[t]{.40\linewidth}
            \includegraphics[width=0.99\linewidth]{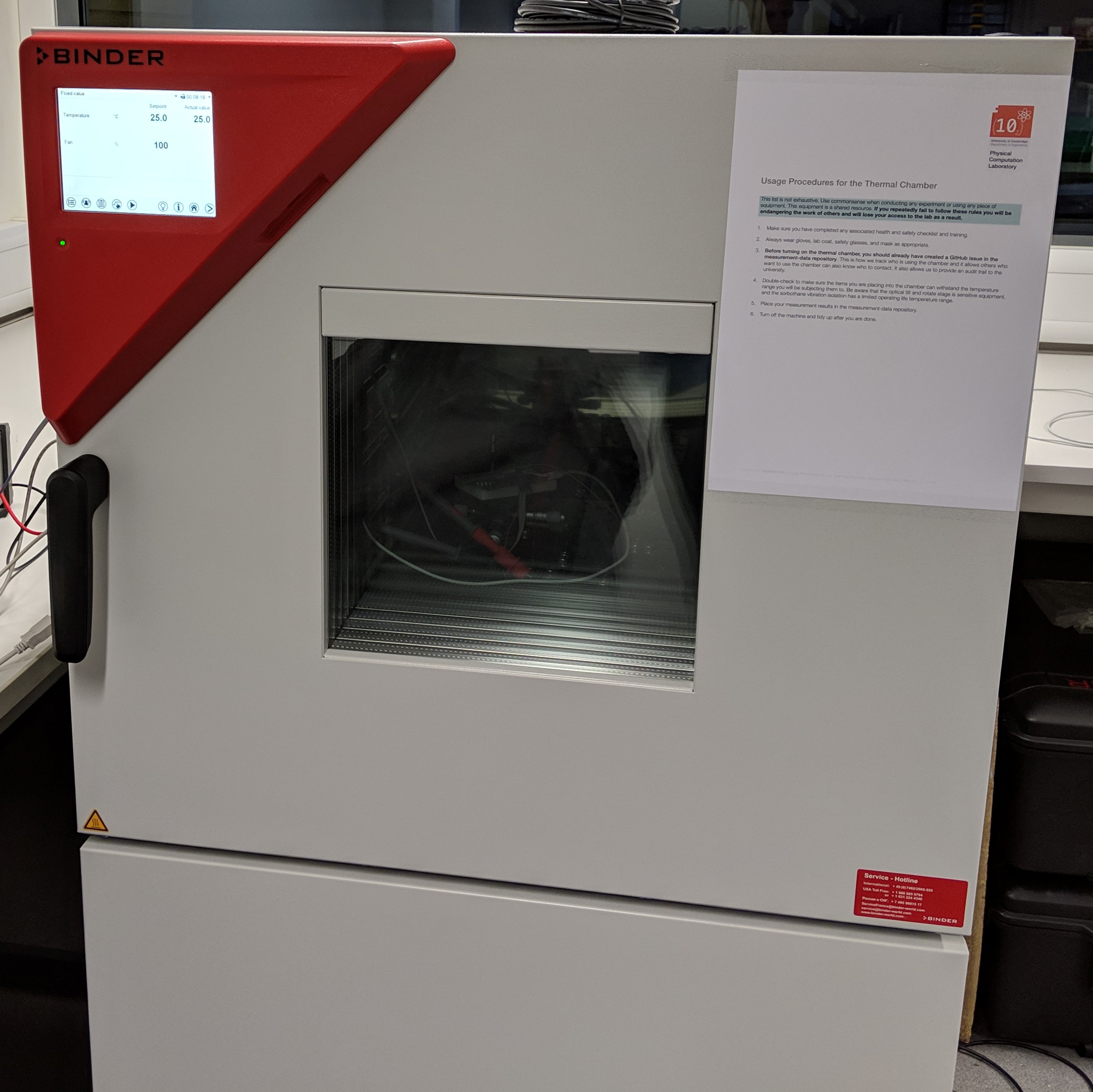}
        \end{subfigure}%
        \begin{subfigure}[t]{.56\linewidth}
            \includegraphics[width=0.99\linewidth]{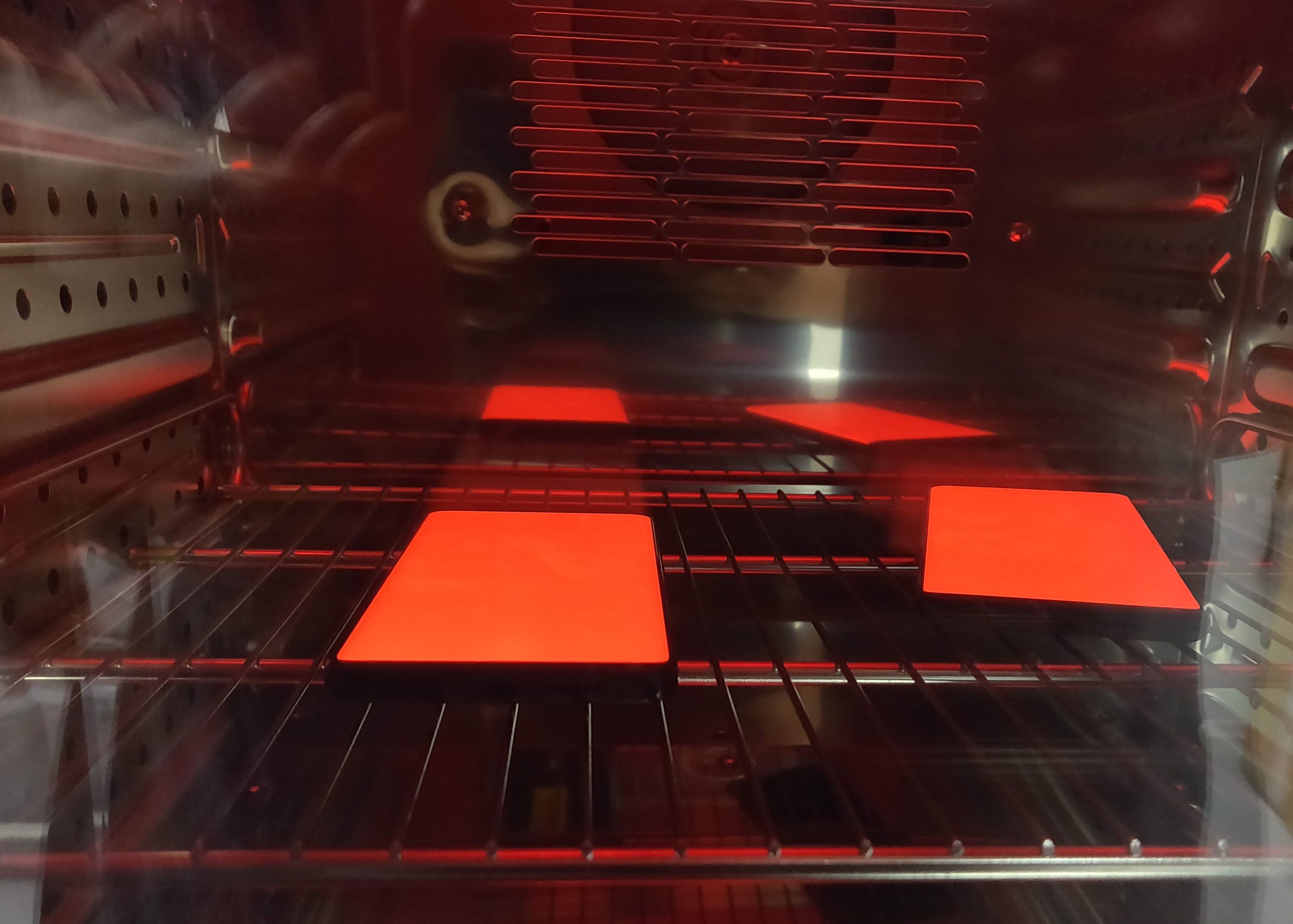}
        \end{subfigure}
        \caption{We used a \textit{Binder MK56} thermal chamber (left) to measure display parameters for the Huawei P30 Pro devices (inside the chamber, right) at 25\textdegree C.}
        \label{fig:dispMeasure}
    \end{minipage}\hfill

\end{figure}

\subsection{Contributions}

We address the challenge of finding the optimal $\lambda$, by using a computationally-efficient model, derived from a user study of acceptability of color-transformed images. The model permits color transformation methods such as Crayon~\cite{Stanley-MarbellMIT2016} to automatically determine their control parameters and to do so intelligently based on the input image. In particular, this work makes the following contributions:

\begin{enumerate}
    \item \textbf{Content-based parameter selection:} It presents an efficient mechanism for calculating color transform parameters based on the image content to achieve a target power saving while preserving a given quality standard (on a mean opinion score (MOS) scale).
    \item \textbf{User study:} A user study with more than 6000 perceptual quality scores from 62 participants, consisting of nearly five times more bitmap images compared to the original Crayon study~\cite{Stanley-MarbellMIT2016}. Results show that we are able to achieve nearly 50\% display power savings with the majority of participants reporting acceptable image quality.
\end{enumerate}

\section{Crayon Color Transform}
\label{section:colorTransform}

At the heart of Crayon is the display power model, which approximates the display power as a function of pixel value as a sum of three quadratic functions~\cite{Stanley-MarbellMIT2016}.
Let $x$ be an $N$-pixel, RGB input image with color channels $r$, $g$, and $b$, and let $x^c[i]$ be the image intensity of channel $c$ at pixel $i$. Additionally, let $\alpha_c$, $\beta_c$, and $\gamma_c$ be the power model parameters for the channel $c\in\{r,g,b\}$.
The total power required to display an image on the display is given by the following equation,
\begin{equation}
    P(x) = \sum\limits_{c \in \{r,g,b\}}\sum\limits^N_{i=1}  \frac{1}{2} \alpha_c x^c[i]^2 + \beta_c x^c[i] + \gamma_c.
    \label{eqn:powerModelForImg}
\end{equation}

Let $y$ be an image that approximates $x$ but causes lower power dissipation on the display. Using a Lagrange multiplier $\lambda$, the operation of finding $y$ can be formulated as the constrained minimization problem below, where $\phi$ is a convex function that measures the distance between the images.
\begin{equation}
    \min\limits_{y} P(y) + \lambda \phi(y-x)
    \label{eqn:constrainedMin}
\end{equation}

\subsection{Least Squares ($\ell_2^2$) solution}

By setting $\phi(y-x) = \frac{1}{2} \| y-x \|_2^2 $ in Equation \ref{eqn:powerModelForImg} and calculating the derivatives with respect to $y^c[i]$, we calculated a global minimizer for $y$. This leads to the solution below.
\begin{equation}
    y^c[i] = \frac{\lambda x^c[i] - \beta_c}{\lambda + \alpha_c}
    \label{eqn:l22soln}
\end{equation}

\subsection{Euclidean-distance ($\ell_2$) solution}

Unlike with the $\ell_2^2$ solution, when $\phi(y-x) = \| y-x \|_2$, the three channels are coupled together. By writing the problem in vector form with $\vec{y_i} = (y^r[i],y^g[i],y^b[i])$, $\vec{x_i}  = (x^r[i],x^g[i],x^b[i])$ and $D_\alpha$ being a diagonal matrix with $\alpha_r$, $\alpha_g$, and $\alpha_b$ on the diagonal and setting $\beta_c = 0$ and $\gamma_c = 0$ to simplify the power model, the problem can be simplified with a change of variable, $\vec{z_i} =\vec{y_i}-\vec{x_i}$ to get
\begin{equation}
    \min\limits_{\vec{z_i}} \frac{1}{2} (\vec{z_i} + \vec{x_i})^T D_\alpha (\vec{z_i} + \vec{x_i}) + \lambda \| \vec{z_i}\|_2
    \label{eqn:l2_changeOfVar}
\end{equation}

This is minimized when $\vec{z_i}$ is opposite to $\vec{x_i}$ (i.e., $\vec{z_i} = -\mu\vec{x_i}$, for some $\mu>0$). Substituting $\vec{z_i} = -\mu\vec{x_i}$ and differentiating with respect to $\mu$ gives the solution below.
\begin{equation}
    \vec{y_i} = (\mu+1)\vec{x_i} \textrm{ with } \mu = max (1 - \frac{\lambda \|\vec{x_i}\|_2}{\vec{x_i}^T D_\alpha \vec{x_i}}, 0)
    \label{eqn:l2soln}
\end{equation}

\section{User Study}
\label{section:userStudy}

The aim of the study was to model the perceptual quality of transformed images to devise a way to dynamically pick optimal values for $\lambda$ in Equations \ref{eqn:l22soln} and \ref{eqn:l2soln}. We used 14 base images from \textit{Unsplash}~\cite{Unsplash2020} for the study (see Figure \ref{fig:imageCollage}).
We used a \textit{Huawei P30 Pro} device as the target platform. We measured its display power as a function of the R,G,B values in a \textit{Binder MK56} thermal chamber with temperature regulated at 25 \textdegree C to remove the effect of ambient temperature on display power consumption (see Figure \ref{fig:dispMeasure}). We used these measurements to derive the power model parameters.
We used, both $\ell_2$ and $\ell_2^2$ distance metrics, and the three color spaces, sRGB, CIE LAB, and CIE UVW, to transform the images, giving 6 distance metric and color space configurations. For each transform configuration, the $\lambda$ limits were chosen such that, when transforming a pure white image, the maximum and minimum $\lambda$ values would produce an image that consumes 95\% and 40\% of the power required to display the original image, respectively. We then normalized the $\lambda$ value to the range $[0,1]$ for each transform configuration.

We transformed the 14 images for each of the 6 configurations using 20 $\lambda$ values, at 5\% intervals in the normalized range. This resulted in 1,680 unique transformed images. We then split the transformed images into batches of 20 images and paired them with their originals. We also added two control image pairs, an identical pair and a pair with one of the images set to a black image, to each batch at random locations to allow us to validate participant responses.

We ran the study on Amazon's \textit{Sage Maker} platform using the \textit{Amazon Mechanical Turk} workforce. We presented each participant with a batch of images and asked them to comment on the visual impairments on the transformed image of each pair, using the following 5-point double-stimulus-impairment-scale~\cite{ITU2012}: Score 5: \textit{Imperceptible}, Score 4: \textit{Perceptible, but not annoying}, Score 3: \textit{Slightly annoying}, Score 2: \textit{Annoying}, Score 1: \textit{Very annoying},

To simulate a real-world observation time, we asked the participants not to spend more than 20 seconds on each image pair. For each batch we collected responses from 5 different participants giving 8,400 image pair comparisons from 62 unique participants.

\begin{figure}[h]
    \centering
    \begin{minipage}{0.45\linewidth}
        \centering
        \includegraphics[width=\linewidth, trim={0.5cm 0cm 2cm 1.6cm}, clip]{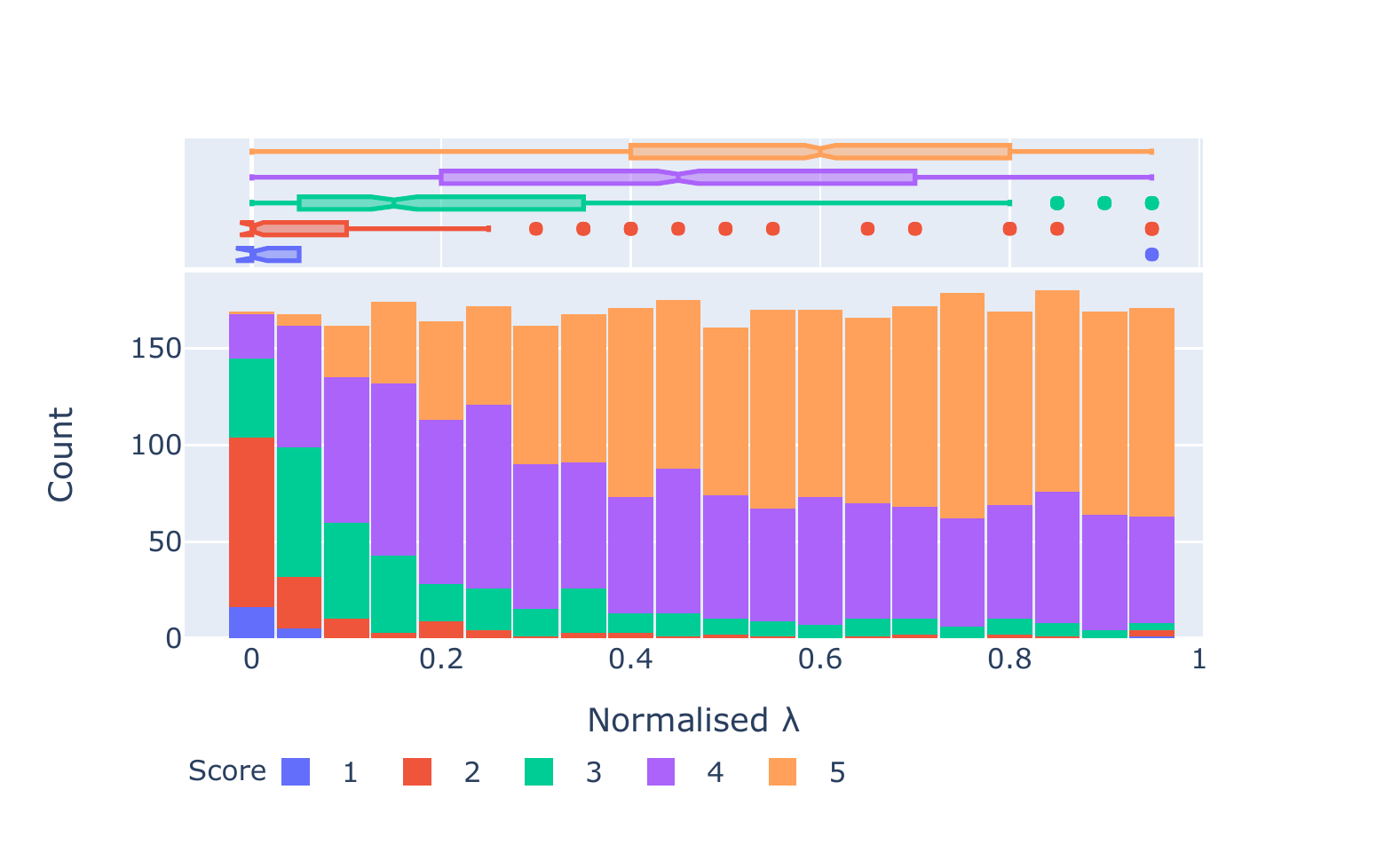}
    \end{minipage}\hfill
    \begin{minipage}{0.45\linewidth}
        \centering
        \includegraphics[width=\linewidth, trim={0.5cm 0cm 2cm 1.6cm}, clip]{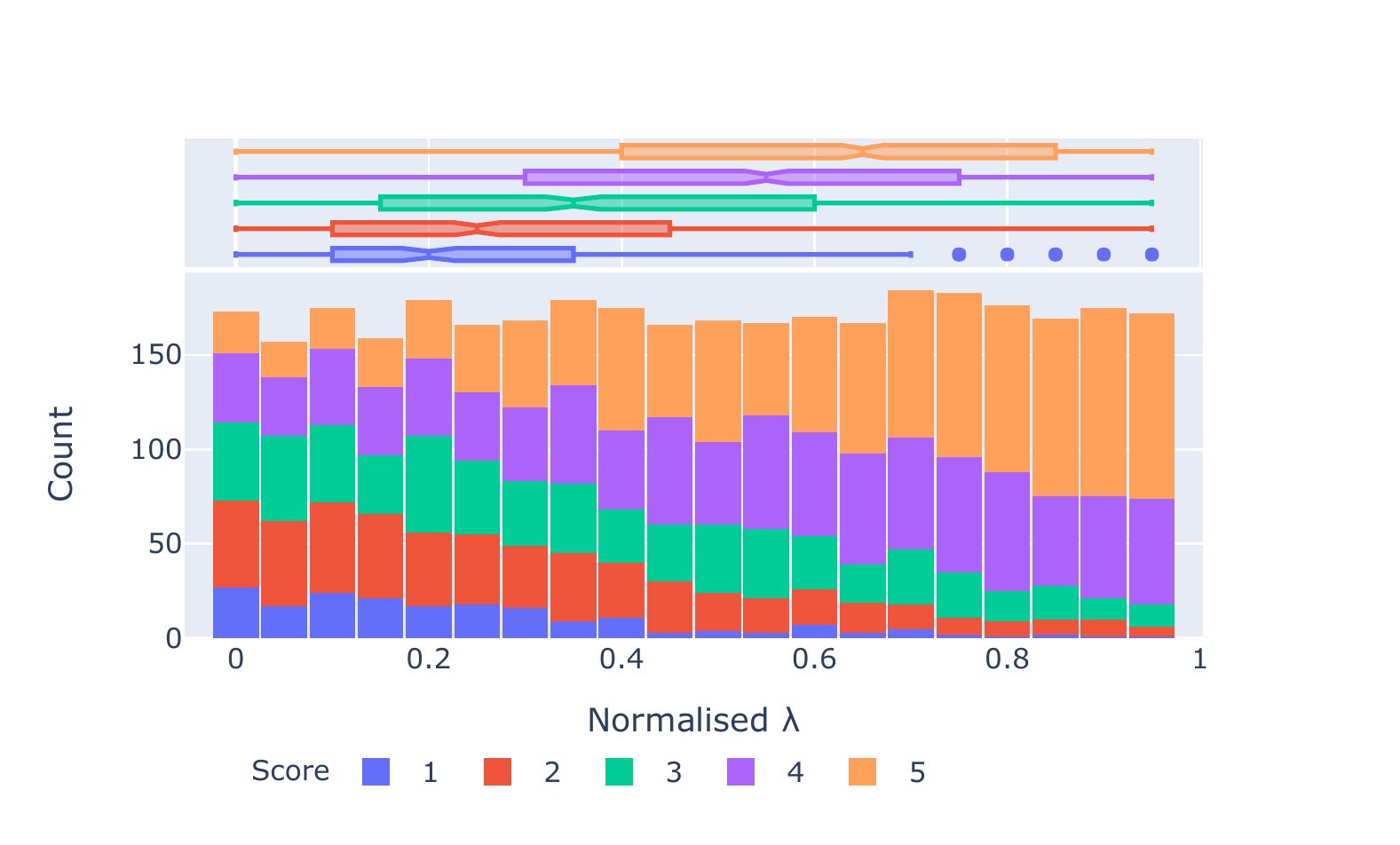}
    \end{minipage}\hfill
    \begin{minipage}{\linewidth}
        \caption{\footnotesize{Raw score histograms show that the perceptual quality is better preserved for a wider range of $\lambda$ when using $\ell_2^2$ distance (top) compared to $\ell_2$ distance (bottom). Box and whisker plots above each histogram show the distribution of the scores over the range of $\lambda$ values on the x-axis.}}
        \label{fig:rawScoreHist}
    \end{minipage}
\end{figure}
First, we discarded any batches that incorrectly labelled either of the control image pairs (i.e., the identical pair or the image paired with a black rectangle), resulting in 6,800 image pairs. Figure \ref{fig:rawScoreHist} shows the distribution of the raw scores as a function of the normalized $\lambda$ value for both $\ell_2^2$ and $\ell_2$ distances. The figure shows that when using the $\ell_2$ distance, the proportion of images with scores 4 and 5 drop off more rapidly than with the $\ell_2^2$ distance, showing that the $\ell_2^2$ distance is better at preserving perceived quality. Even down to a $\lambda$ value of 0.1 (\textasciitilde50\% power saving), the majority of the participants said that the artifacts were \textit{"Imperceptible"} or \textit{"Perceptible, but not annoying"} for images that were transformed using $\ell_2^2$ distance. These observations on the difference in the effect of $\ell_2$ and $\ell_2^2$ were also reported in the original Crayon work. The results we present however encompass about five times as many unique images compared to the original crayon study.

\begin{figure}
    \begin{minipage}{0.49\linewidth}
        \centering
        \includegraphics[width=\linewidth, trim={0.5cm 1.8cm 2cm 2.5cm}, clip]{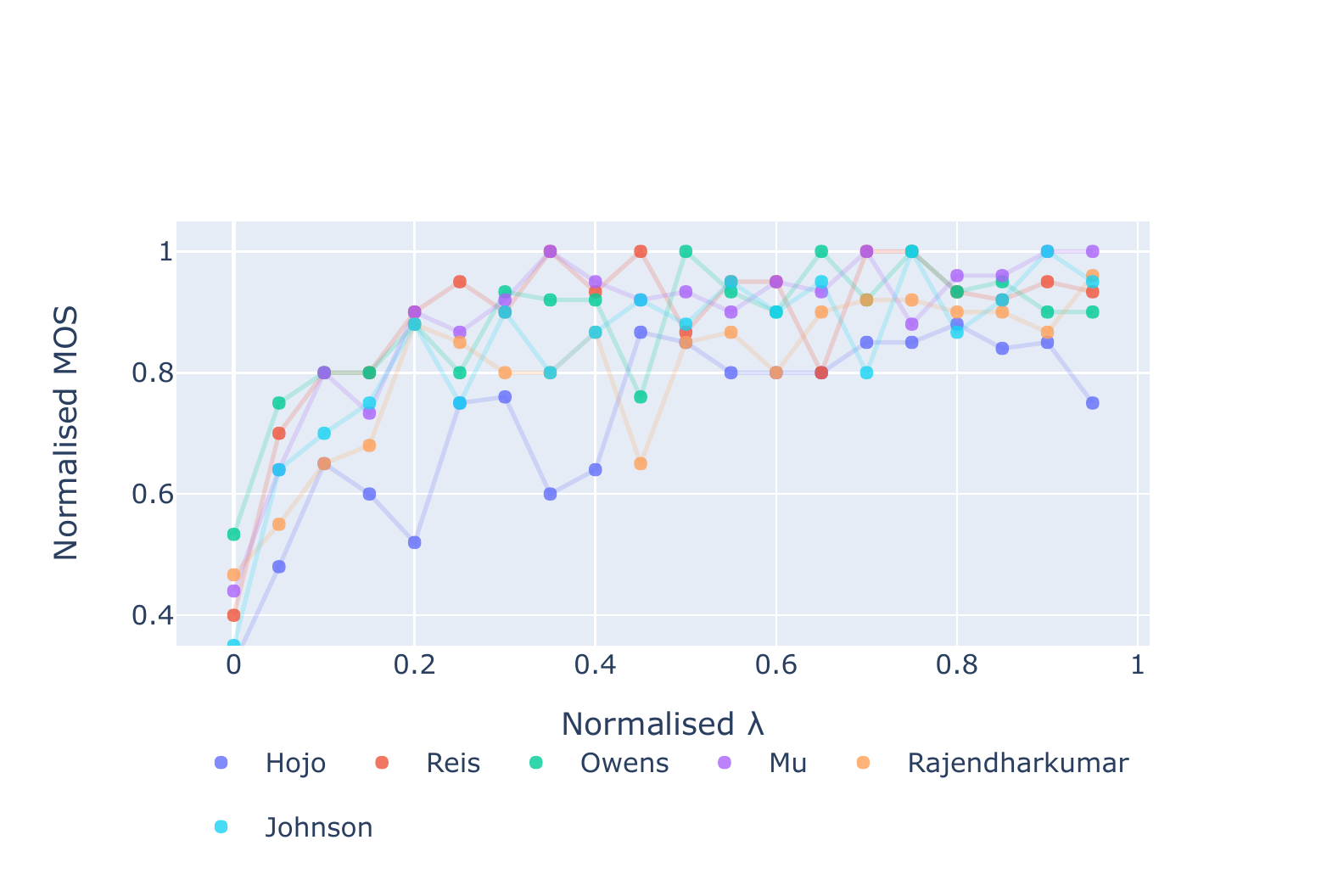}
    \end{minipage}\hfill
    \begin{minipage}{0.49\linewidth}
        \centering
        \includegraphics[width=\linewidth, trim={0.5cm 1.8cm 2cm 2.5cm}, clip]{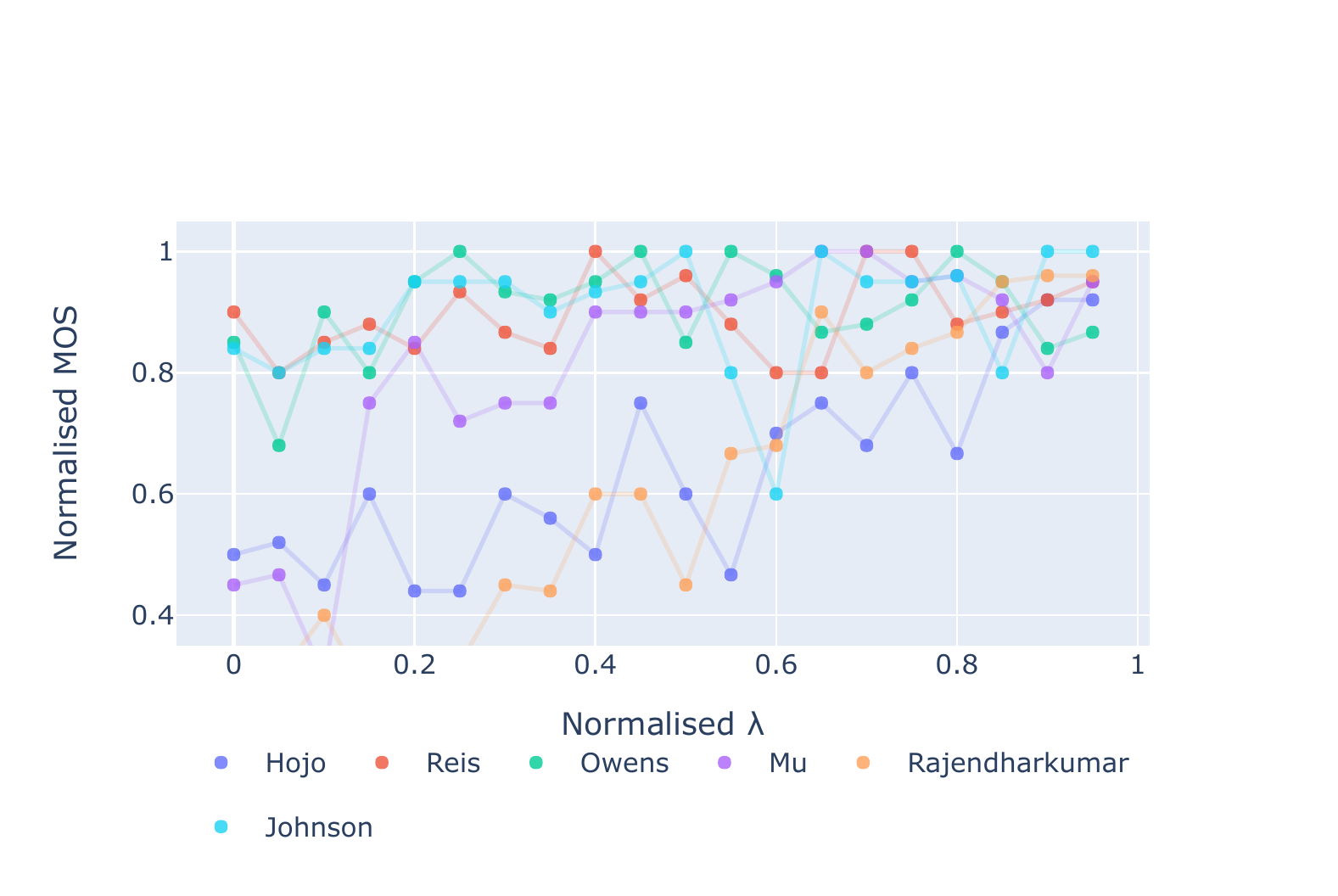}
    \end{minipage}
    \begin{minipage}{1\linewidth}
        \centering
        \includegraphics[width=0.5\linewidth, trim={0 0 0.5cm 0}, clip]{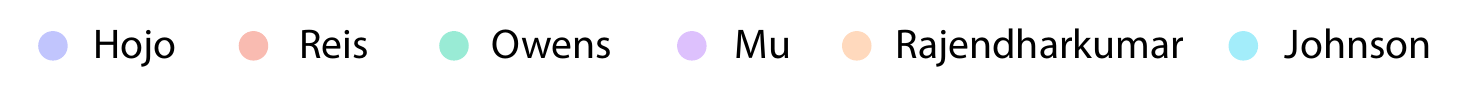}
        \caption{MOS variation with $\lambda$ for different images follow a similar trend when using $\ell_2^2$ distance (top) in contrast to $\ell_2$ distance (bottom).} %
        \label{fig:mosVlambda}
    \end{minipage}
\end{figure}

\subsection{Results}

The individual responses for each unique image pair were then averaged to get mean opinion scores (MOS).
Figure \ref{fig:mosVlambda} shows the normalized MOS against $\lambda$ for the images in Figure \ref{fig:egImages}.
It shows that the change in MOS follows a similar trend for all the images when using $\ell_2^2$ distance while the results for the images using $\ell_2$ distance does not show as much similarity.
We used this trend similarity when using $\ell_2^2$ distance, to formulate a method for efficiently predicting the expected user response, based just on the input image.
Furthermore, the normalization of $\lambda$ and MOS will allow the results to be compared between different transform settings and be used with different power models.

\begin{figure}[h]
    \centering
    \begin{minipage}{0.34\linewidth}
        \centering
        \includegraphics[width=\linewidth, trim={0.5cm 2cm 2cm 2cm}, clip]{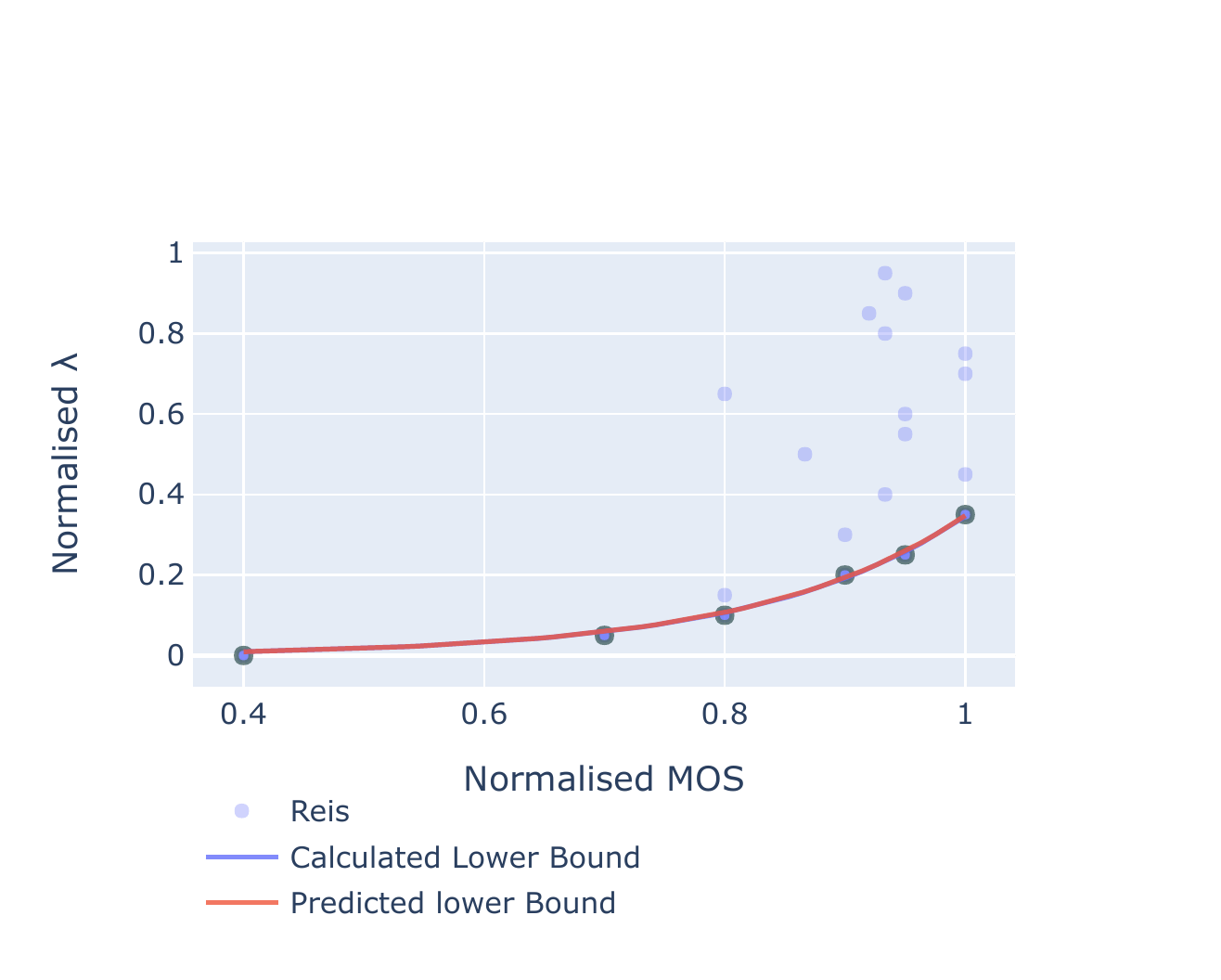}

    \end{minipage}
    \begin{minipage}{0.32\linewidth}
        \centering
        \includegraphics[width=\linewidth, trim={1.5cm 2cm 2cm 2cm}, clip]{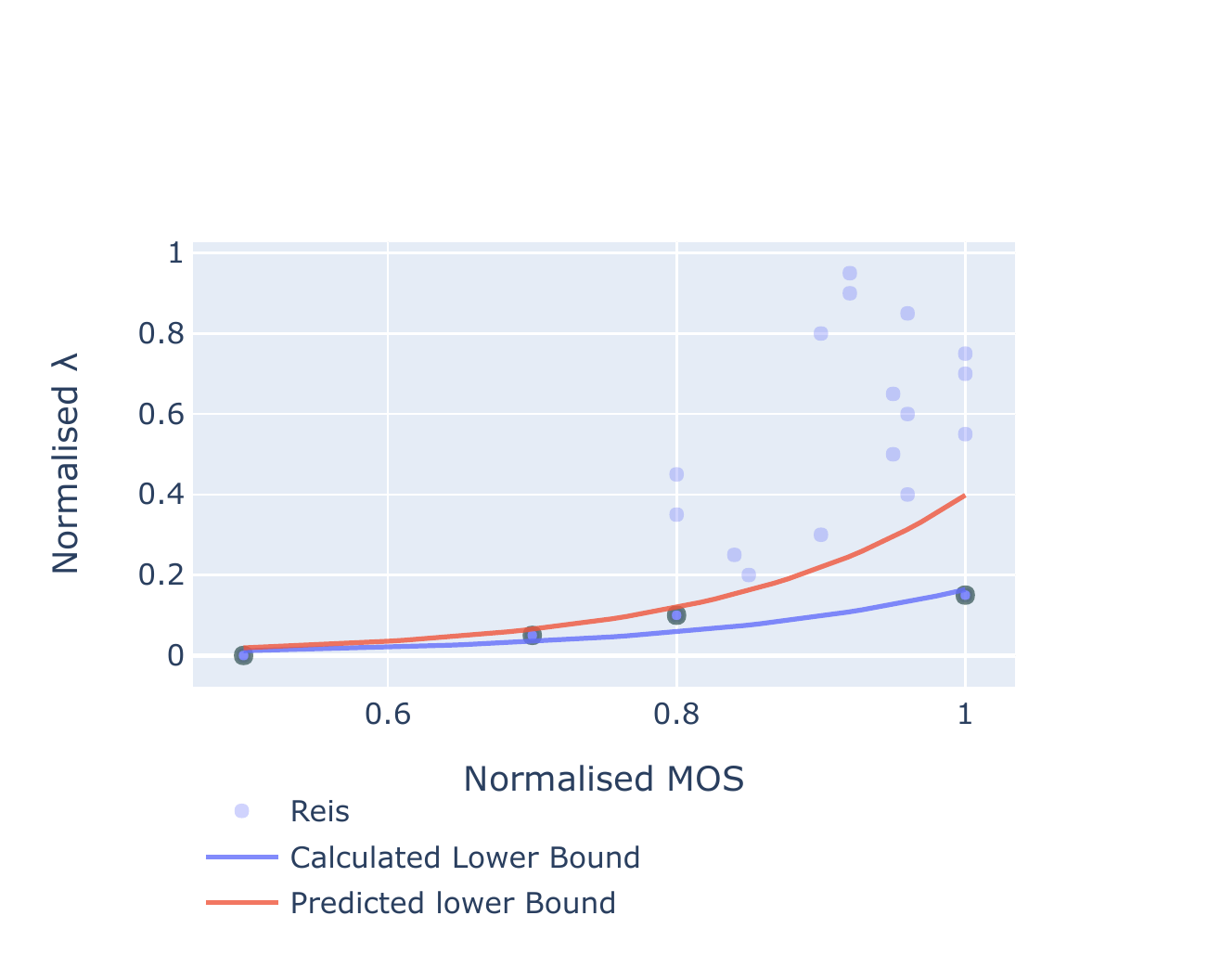}
    \end{minipage}
    \begin{minipage}{0.32\linewidth}
        \centering
        \includegraphics[width=\linewidth, trim={1.5cm 2cm 2cm 2cm}, clip]{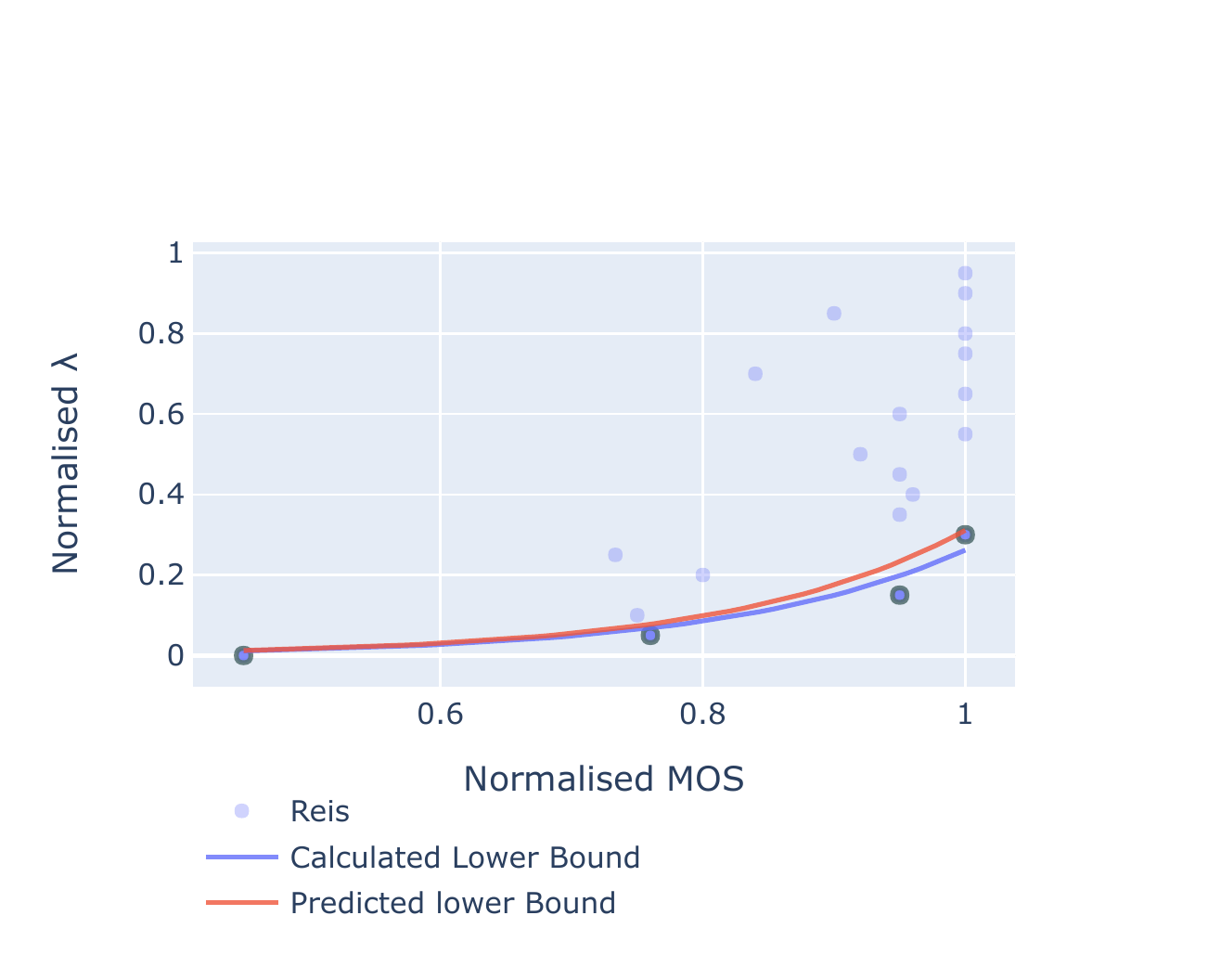}
    \end{minipage}
    \begin{minipage}{0.4\linewidth}
        \centering
        \includegraphics[width=\linewidth, trim={0 0.2cm 0.5cm 0.4cm}, clip]{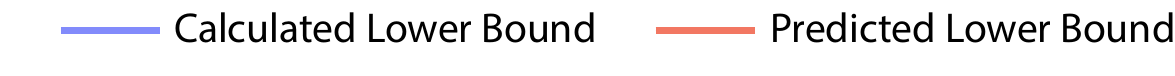}
    \end{minipage}
    \begin{minipage}{1\linewidth}
        \centering
        \caption{\footnotesize{Predicted lower bounds for the image in Figure \ref{fig:reis} using the \textit{leave-out} SVM model follow the trend of the data points better than the calculated fit for all three color spaces; RGB (top), CIE LAB (middle), and CIE UVW (bottom).}}
        \label{fig:lbEstCurves_predicted_reis}
    \end{minipage}%
\end{figure}

\begin{figure}
    \centering
    \begin{minipage}{0.34\linewidth}
        \centering
        \includegraphics[width=\linewidth, trim={0.5cm 2.5cm 2cm 2.8cm}, clip]{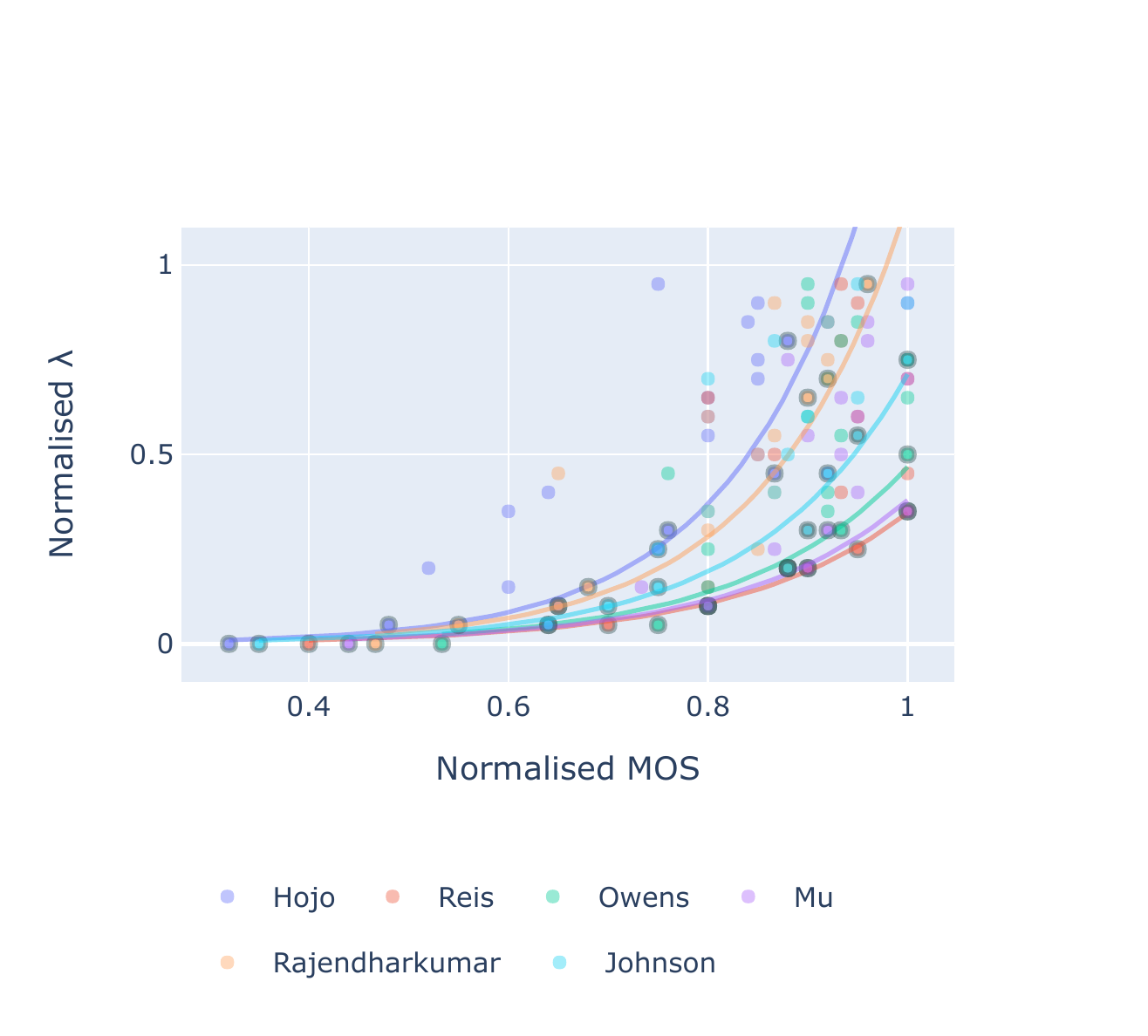}
    \end{minipage}
    \begin{minipage}{0.32\linewidth}
        \centering
        \includegraphics[width=\linewidth, trim={1.3cm 2.5cm 2cm 2.8cm}, clip]{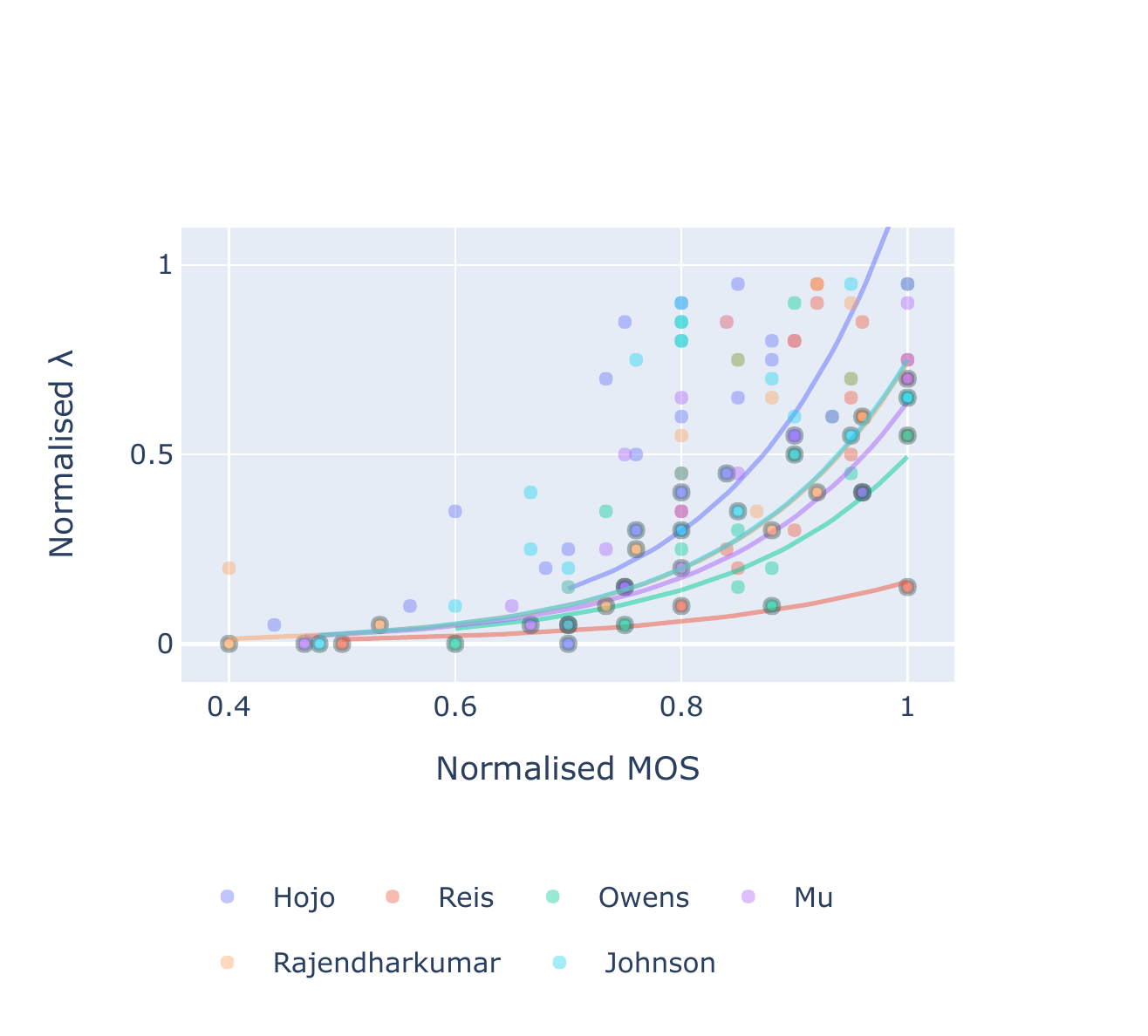}
    \end{minipage}
    \begin{minipage}{0.32\linewidth}
        \centering
        \includegraphics[width=\linewidth, trim={1.3cm 2.5cm 2cm 2.8cm}, clip]{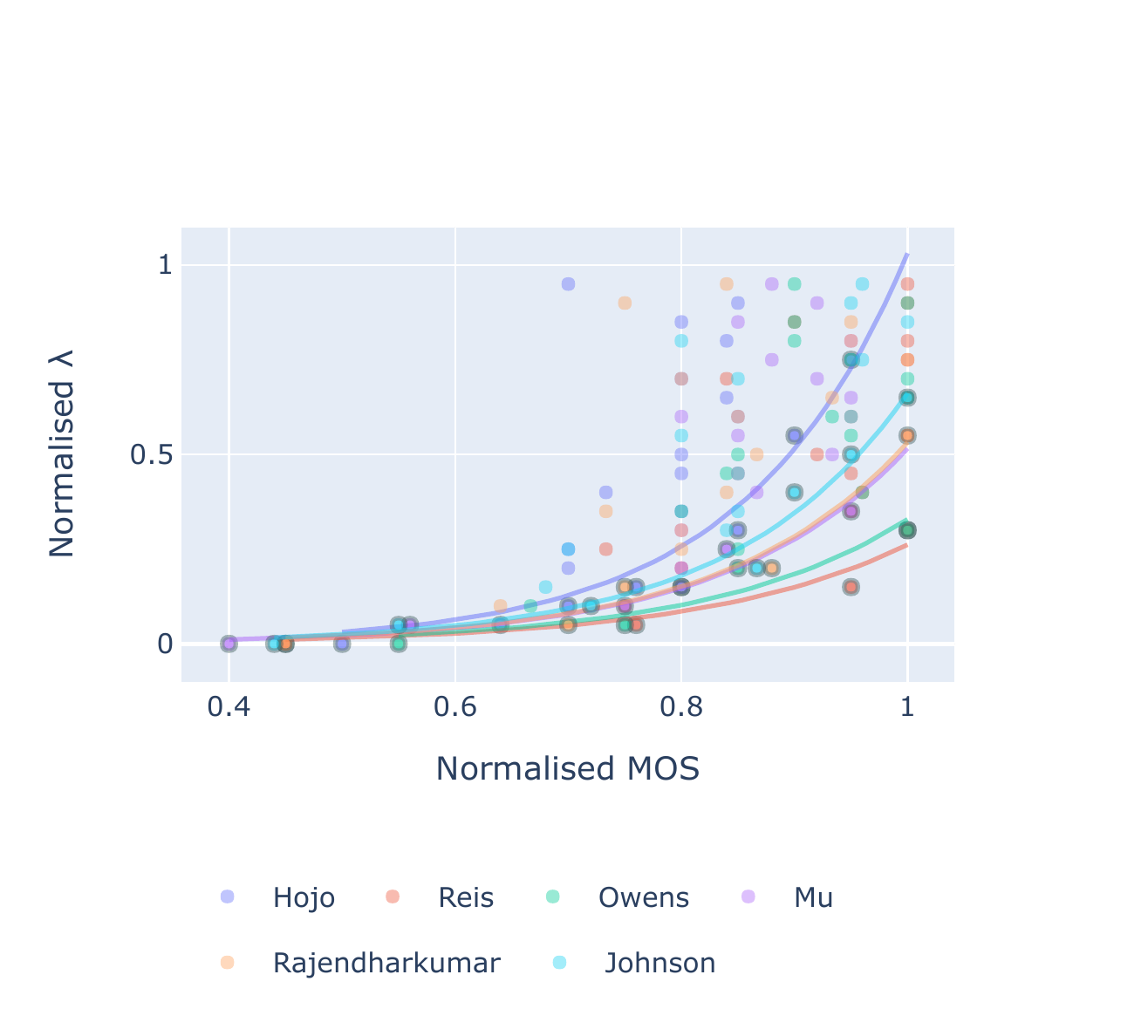}
    \end{minipage}
    \begin{minipage}{1\linewidth}
        \centering
        \includegraphics[width=0.5\linewidth, trim={0 0 0.5cm 0}, clip]{Illustrations/lambdVscore_lbFit_legend.pdf}
        \caption{\footnotesize{The exponential model can fit the lower bound for $\lambda$ against MOS when using $\ell_2^2$ distance for all three color spaces; RGB (top), CIE LAB (middle), and CIE UVW (bottom).}}
        \label{fig:lbEstCurves}
    \end{minipage}
\end{figure}
\subsection{Lambda ($\lambda$) Lower Bound Prediction}

When using Crayon in a system, we need to decide on $\lambda$ based on what MOS we require.
In the study, we set $\lambda$, and we got a range of scores from the users.
However, the mapping from $\lambda$ to scores is multi-valued, and thus it is not invertible.
Therefore, we chose to find what the smallest value of $\lambda$ that gives a particular MOS would be, to derive a function for the lower bound for $\lambda$ given MOS.
To find an analytic solution for the lower bound, we filtered the results to find the scores on the lower boundary and fit the exponential below to these data points by adjusting the rate parameter $k$.
\begin{equation}
    \lambda_{\textnormal{\footnotesize{LB}}}(s)=\frac{e^{k s}-1}{1000}
    \label{eqn:lbExp}
\end{equation}

Figure \ref{fig:lbEstCurves_predicted_reis} shows the raw $\lambda$ values plotted against MOS for image in Figure \ref{fig:reis}. The figure shows the data points on the lower bound highlighted in bold and the blue line is the exponential fit. Figure \ref{fig:lbEstCurves} shows the result of this process for the six images in Figure \ref{fig:egImages}.

The aforementioned process for finding the lower bound relies on user study data and consequently, it is only applicable to the 14 images used in the study. In practice, we need to find the lower bound for a previously-unseen image. So we need to generalize the results from the user study to find a correlation between image properties and the lower bound.

We investigated 12 simple heuristics based on hue, luminance, and saturation distribution of the image pixels, to generalize beyond the data points from the user study. The aim was to see if $k$ is correlated with any of these.
For the original images used in the study, we found that mean luminance, and standard deviation of the saturation and hue across the image were correlated with the parameter $k$ with Pearson and Spearman rank correlation coefficients around 0.7 (p<0.01).

Based on this we chose to use mean luminance, and the standard deviation of hue, saturation, and luminance as the image features to use in predicting $k$ (i.e., the shape of the exponential).

\begin{table}[h]
    \centering
    \begin{minipage}{0.5\linewidth}
        \begin{small}
            \begin{tabular}{ l  c c c r }
                \textbf{Color Space} & \textbf{Model}  & \textbf{MSE}     & \textbf{Variance} & \textbf{\% error} \\
                \hline
                                     & Linear          & 0.74988          & 0.36193           & 38.325            \\
                RGB                  & Cubic           & 0.57005          & 0.19253           & 29.134            \\
                                     & \textbf{SVM}    & \textbf{0.29892} & \textbf{0.07820}  & \textbf{15.277}   \\
                                     & SVM             & 0.17525          & 0.01067           & 8.6824            \\ \hline
                                     & \textbf{Linear} & \textbf{0.14359} & \textbf{0.00469}  & \textbf{7.114}    \\
                CIE LAB              & Cubic           & 1.82583          & 4.45206           & 90.457            \\
                                     & SVM             & 0.17525          & 0.01067           & 8.6824            \\ \hline
                                     & \textbf{Linear} & \textbf{0.18351} & \textbf{0.01352}  & \textbf{9.955}    \\
                CIE UVW              & Cubic           & 0.36907          & 0.06915           & 20.020            \\
                                     & SVM             & 0.28552          & 0.01677           & 15.488            \\
                \hline
            \end{tabular}
        \end{small}
        \caption{Our model is able calculate the lower bound parameter $k$ to a mean percentage MSE of 7\%, with 5-fold cross validation.}
        \label{tab:kModelResults}
    \end{minipage}\hfill
    \begin{minipage}{0.45\linewidth}
        \centering
        \small{
            \begin{tabular}{ l  c c r }
                \textbf{Color Space} & \textbf{Model} & \textbf{MSE} & \textbf{\% error} \\
                \hline
                RGB                  & SVM            & 0.08008      & 4.091             \\
                CIE LAB              & SVM            & 0.17532      & 8.686             \\
                CIE UVW              & SVM            & 0.02951      & 1.601             \\
                \hline
            \end{tabular}
        }
        \caption{When using 13 of the 14 original images, our SVM model is able to further increase its accuracy to 1.6\% showing that the accuracy could be increased with more data.}~\label{tab:loutModelResults}
    \end{minipage}
\end{table}

Table \ref{tab:kModelResults} summarizes the average mean squared error (MSE) and MSE variance achieved for each color space, with 5-fold cross validation, by linear, cubic, and support vector machine (SVM) models. The MSE is the error between the $k$ value predicted using the image heuristics and the value calculated with the user study results. We used the range of calculated $k$ values for each color space was to calculate the percentage MSE.

We are able to predict the parameter $k$ to a maximum mean accuracy, over the 5-folds, of about 7\% MSE. For both CIE LAB and CIE UVW color spaces, a simple linear model performs the best, while in RGB color space the SVM model has the best performance. We trained a further set of SVM models using all the images except the image in Figure \ref{fig:reis}. Table \ref{tab:loutModelResults} shows the results for these models and additionally, Figure \ref{fig:lbEstCurves_predicted_reis} shows the lower bound prediction. These models performed with a maximum accuracy of 1.6\% MSE when used to predict the lower bound for the image in Figure \ref{fig:reis}. This indicates that the accuracy of our model could be increased with more data. This is an area we are actively exploring as an extension to this "late breaking results" submission. Additionally the model is able to avoid fitting to the outlier point at MOS = 1 in the CIE LAB color space (Figure \ref{fig:lbEstCurves_predicted_reis}, middle). This indicates the learned model has generalized well.

This work can be extended with a larger user study with a larger set of images and more flexible models to increase the accuracy of the predictions.
However, this carries the risk of getting to unreasonably numerical high accuracies and a false sense of safety. Given the inherent variation in subjective opinion, even within a single person, high numerical accuracies would result in a misplaced confidence in the real world performance of a given model. However, a larger study would also allow us to calculate useful, mean and variance estimates for user score. This would ultimately let our system provide accurate confidence intervals for the proportion of people that would score the image to be of satisfactory quality, given the chosen $\lambda$.

\section{Related Research}
\label{section:relatedwork}

Crayon is the work most closely related to this work~\cite{Stanley-MarbellMIT2016}. But as outlined before, the original work does not provide a way to automatically tune transform parameters, which our work addresses. Stanley-Marbell et al. have proposed a similar system called Ishihara, that uses a large corpus of color matching data to merge easily confused colors to improve efficiency~\cite{Stanley-Marbell2018b}.
Most of the work on OLED power optimization in literature focusses on GUI color remapping~\cite{Dong2009}, replacing UI colors with black~\cite{Li2014} or obfuscating part of the screen~\cite{Wee2013}. Our work could be integrated with Chameleon~\cite{Dong2009} as it uses a similar formulation to Crayon~\cite{Stanley-MarbellMIT2016}. Previous work exploring color mapping to minimize the use of the blue subpixels~\cite{Anand2014}, display vignetting to reduce brightness in display edges~\cite{Wee2012}, or brightness scaling to reduce overall brightness of the display~\cite{JahierPagliari2019} exist in literature.

Our work can also be posed as a no-reference (NR) image quality metric (IQM) for color approximation algorithms. The calculated $\lambda$ lower bound function could be inverted to infer MOS from image and $\lambda$. Typical NR-IQMs rely on a human vision model to infer the effects of artifacts and focus on luminance information in the image~\cite{Ferzli2009,Liu2014}. Although they perform well for images with luminance artifacts, their performance on color artifacts have not been explored specifically.
Other possible extensions include combining the color transformations with power-saving I/O encoding techniques~\cite{2016:RSI:2897937.2898079} or even inferring permissible color approximation from programming languages that permit programmers to specify accuracy constraints~\cite{M:pmup06}.

\section{Acknowledgements}
This research is supported by an Alan Turing Institute award TU/B/000096
under EPSRC grant EP/N510129/1. C. Samarakoon is supported by the EPSRC DTP Studentship award EP/N509620/1. The images used in the study are reproduced under the Unsplash Creative Commons license (CC0). The original authors of the images are
Abigail Mangum,
Nirmal Rajendharkumar,
Simone Hutsch,
Jakob Owens,
Alex Meier,
Steve Johnson,
Jon Tyson,
Akira Hojo,
Amber Kipp,
Mae Mu,
Alexander Schimmeck,
Alexander Akimenko,
Andre Reis and
Zhenzhong Liu,
left to right, top to bottom, as shown in Figure~\ref{fig:imageCollage}.

\bibliography{crayon-refs}
\bibliographystyle{ACM-Reference-Format}

\end{document}